\documentclass[usenatbib,usegraphicx]{mn2e}

\def\ApJ{ApJ}
\def\ApJS{ApJS}
\def\ApJL{ApJL}
\def\AJ{AJ}
\def\PRD{Phys. Rev. D}
\def\MNRAS{{\it Mon. Not. Roy. Astron. Soc.}}

\def\JCAP{{\it JCAP}}

\def\araa{ARA\&A}

\begin{document}

\title{The SDSS DR7 Galaxy Angular Power Spectrum: Volume-Limits and Galaxy Morphology}
\author[B. Hayes et al.]{Brett Hayes$^1$ and Robert Brunner$^1$ \\ $^1$Department of Astronomy, University of Illinois, Urbana, IL}

\maketitle
\begin{abstract}
We use a quadratic estimator with KL-compression to calculate the angular power spectrum of a volume-limited Sloan Digital Sky Survey (SDSS) Data Release 7 (DR7) galaxy sample out to $\ell = 200$.  We also 
determine the angular power spectrum of selected subsamples with photometric redshifts $z < 0.3$ and $0.3 < z <0.4$ to examine the possible evolution of the angular power spectrum, as well as early-type and 
late-type galaxy subsamples to examine the relative linear bias.  In addition, we calculate the angular power spectrum of the SDSS DR7 main galaxy sample in a $\sim 53.7$ square degree area out to $l = 1600$ 
to determine the SDSS DR7 angular power spectrum to high multipoles.  We perform a $\chi^2$ fit to compare the resulting angular power spectra to theoretical nonlinear angular power spectra to extract 
cosmological parameters and the linear bias.  We find the best-fit cosmological parameters of $\Omega_m = 0.267 \pm 0.038$ and $\Omega_b = 0.045 \pm 0.012$.  We find an overall linear bias of 
$b = 1.075 \pm 0.056$, an early-type bias of $b_{e} = 1.727 \pm 0.065$, and a late-type bias of $b_{l} = 1.256 \pm 0.051$.  Finally, we present evidence of a selective misclassification of late-type galaxies 
as stars by the SDSS photometric data reduction pipeline in areas of high stellar density (\textit{e.g.}, at low Galactic latitudes).
\end{abstract}
\begin{keywords}
galaxies: statistics -- large-scale structure of the universe -- methods: data analysis
\end{keywords}

\section{Introduction} 
\label{Introduction}

The amount of normal, baryonic matter, dark matter, and dark energy in the Universe controls the formation and evolution of large scale structure.  Hence, by observing the distribution of the large 
scale structure, as traced by the galaxy density, we can determine the densities of these components of the Universe.  The distribution of galaxies can be statistically quantified by the three dimensional 
power spectrum (see, \textit{e.g.}, \citealt{percival07}).  Spectroscopic surveys, however are generally limited by either sample size, cosmic volume, or both, as obtaining spectroscopic redshifts for 
a large sample of faint galaxies can be expensive and difficult.  Instead, we can utilize the large photometric surveys of galaxies to more precisely measure the angular power spectrum 
(\citealt{tegmark02}), and compare these measurements to a theoretical 3D power spectrum by using the photometric redshift distribution of the galaxy sample.

From these comparisons, we can extract constraints on any parameters that effect the power spectrum, including bias (\citealt{blake04}), cosmological matter density, $\Omega_m$ (\citealt{huterer01}), 
baryon density, $\Omega_b$ (\citealt{frith05}), or even neutrino masses (\citealt{deputter12}).  In addition to the parameter space explored by galaxy angular power spectra, the natural relationship 
between the angular power spectrum and the 3D power spectrum can be used to reconstruct the 3D power spectrum (\textit{e.g.}, \citealt{dodelson02}).  Due to these useful properties, galaxy angular power 
spectra have been estimated for many surveys, including the SDSS (\citealt{tegmark02,thomas10,ho12}), EDSGC (\citealt{huterer01}), NVSS (\citealt{blake04}), 2MASS (\citealt{frith05}), and others 
(\textit{e.g.}, \citealt{baugh94}).  

In a previous paper (\citealt{hayes12};  hereafter HBR12), we detailed our implementation of the \cite{bond98} quadratic 
estimator to calculate the angular power spectrum of the SDSS DR7 using HEALPix.  With this technique, we used our calculated angular power spectra together with a projected linear 3D power spectrum to 
retrieve constraints on cosmological matter density and bias.  The cosmological parameters of greatest interest, however, impact the angular power spectrum at multipoles that correspond to small scales, 
that is in the nonlinear regime.  To more tightly constrain cosmological parameters, it is necessary to fit estimated angular power spectra to theoretical, nonlinear 3D matter power spectra, for example, 
by using the Code for Anisotropies in the Microwave Background, also known as CAMB (\citealt{lewis00}).

In this paper, therefore, we build on this previous work to present new angular power spectrum measurements.  However, we have chosen first to calculate angular power spectra of galaxies from the SDSS 
DR7 using a volume-limited sample created by \cite{ross10}.  This allows us to make subsamples that can be directly compared to each other, such as high and low redshift samples to 
investigate the evolution of the angular power spectrum, as well as galaxy-type defined samples that allow for a measurement of the relative linear bias between early- and late-type galaxies.  This 
current paper first begins with an overview of our data, quadratic estimator, and $\chi^2$ fitting process in Section \ref{Overview}.  Next, we use CAMB to determine theoretical angular power spectra 
to compare with the estimated SDSS DR7 angular power spectra and detail the effects this has on the $\chi^2$ fits to extract cosmological parameters.  In Section \ref{Volume-Limited Sample}, we 
determine the galaxy angular power spectrum using a volume-limited sample of SDSS DR7 galaxies that is further divided into two redshift slices to quantify the redshift evolution of our quantified 
cosmological parameters and the linear bias.  We split this volume-limited sample into early- and late-types in Section \ref{Galaxy Morphology}, and perform the same analysis to examine the relative 
bias between galaxy types, as well as combine the type cuts and redshift slices to look at the evolution of the relative bias.  Finally, we examine the angular power spectrum of SDSS DR7 galaxies at 
high multipoles, by constraining our analysis to a subset of the full DR7 and applying our quadratic estimator to a higher resolution map in Section \ref{Large Multipoles}.  We discuss the results in 
Section \ref{Discussion} and conclude in Section \ref{Conclusions}.

\section{Overview}
\label{Overview}

The basic data, systematic cuts, and angular power spectrum estimation technique are fully detailed in HBR12.  Here we provide a concise overview to explain the basics of our analysis.

\subsection{SDSS DR7 Data}
\label{SDSS DR7 Data}

We have chosen a contiguous area of the SDSS DR7 North Galactic Cap ellipsoid to examine, using SDSS stripes 9--37.  Our SDSS DR7 data consists of all galaxies in this area with apparent r-band magnitudes from 
18--21 with photometric redshifts and photometric redshift errors.  We have created subsamples of this data for our analysis, a volume-limited sample and associated subsamples described and used in Sections 
\ref{Volume-Limited Sample} and \ref{Galaxy Morphology}, and a selected area of the main galaxy sample around the North Galactic Pole described in Section \ref{Large Multipoles}.

These samples are pixelated with HEALPix and we mask pixels that fully or partially lie outside of the SDSS DR7 footprint, have average seeing greater than 1.5 arcseconds, have average reddening greater than 
0.2 magnitudes, and have areas of poor observing quality.  Furthermore, any pixel that has less than 75\% usable area after masking is also rejected.  After pixelating with HEALPix and masking, we create a 
vector of pixelized overdensities:

\begin{equation}
x_i \equiv \frac{G_i}{\overline{G}\Omega_i} - 1
\end{equation}
where $G_i$ is the number of galaxies in pixel $i$, $\overline{G}$ is the survey average galaxy density, and $\Omega_i$ is the pixel area.

As the SDSS is not a full sky survey, we are prevented from determining each multipole moment individually.  Multipole resolution is determined by $\Delta \ell \approx 180^{\circ} / \phi$, where $\phi$ is the 
smallest angular dimension of the survey (\citealt{peebles80}).  By using the full area of the SDSS, we can achieve a multipole resolution of $\Delta \ell \approx 5$, and thus we group multipole 
moments together into bands.  Since we cannot determine the power of individual multipole moments, we have assumed that multipole moments in a band have the same value, which also simplifies the computational 
complexity of this quadratic estimation method.  Since we do not expect large variations between adjacent multipoles in the galaxy angular power spectrum, as evidenced by the smooth theoretical predictions, 
this assumption is both justified and convenient.

\subsection{Quadratic Estimation}
\label{Quadratic Estimation}

For our implementation of the quadratic estimation technique, we first construct a covariance matrix from our pixelixed overdensities, and we want to compare this to a model covariance matrix composed of a 
signal matrix plus a noise matrix:

\begin{equation}
\label{EQ covariance matrix}
C_{ij} \equiv \langle x_i x_j \rangle = S_{ij} + N_{ij} ,
\end{equation}
where the signal matrix describes the covariance matrix we expect based on an initial angular power spectrum and the geometry of our survey area, and the noise matrix is modeled as a Gaussian random process:

\begin{equation}
\label{EQ Signal Matrix}
S_{ij} = \sum_{\ell} \frac{2\ell+1}{2\ell(\ell+1)} \mathcal{C}_\ell P_\ell (\cos \theta_{ij}) e^{-\ell(\ell+1)\tau^2}
\end{equation}

\begin{equation}
N_{ij} = \sigma_i^2 \delta_{ij} = \frac{1}{\overline{G}}\delta_{ij} ,
\end{equation}
where the $P_\ell$ are the Legendre polynomials, $\theta_{ij}$ is the angle between pixels $i$ and $j$, $\tau$ is the beam width, and $\sigma_i$ is the rms noise in pixel $i$.  Using the assumption that the 
$\mathcal{C}_\ell$ in a band are equal, we define the $\textbf{P}_b$ matrices for each band:

\begin{equation}
P_{ij} = \sum_{\ell \in b} \frac{2\ell+1}{2\ell(\ell+1)} P_\ell (\cos \theta_{ij}) e^{-\ell(\ell+1)\tau^2}
\end{equation}

Our goal is to determine the $\mathcal{C}_\ell$ that produce a model covariance matrix that matches the covariance matrix constructed from our pixelized data.  We have implemented the quadratic estimator of 
\cite{bond98} with KL-compression (\citealt{vogeley96,tegmark97b}):

\begin{equation}
\label{estimator}
\delta \mathcal{C}_b = \frac{1}{2} (\textbf{F}^{-1/2})_{b b'}\ \textrm{Tr}\left[(\textbf{x}' \textbf{x}'^T-\textbf{N}')(\textbf{C}'^{-1}
\textbf{P}'_{b'} \textbf{C}'^{-1})\right]
\end{equation}
where, the Fisher information matrix \textbf{F} is defined as:

\begin{equation}
\label{fisher}
F_{b b'} = \frac{1}{2}\ \textrm{Tr}\left(\textbf{C}'^{-1} \textbf{P}'_{b} \textbf{C}'^{-1} \textbf{P}'_{b'}\right) .
\end{equation}

We use $\textbf{F}^{-1/2}$ for this estimator as suggested by \cite{tegmark98} to provide uncorrelated bandpower error bars, given by $\sigma_b = \sqrt{(\textbf{F}^{-1})_{bb}}$, and well behaved window 
functions.  This computationally demanding process can be iterated and converges quickly given a reasonable input power spectrum.

After determining the angular power spectrum, we can recombine adjacent bandpowers together to produce higher signal-to-noise measurements at the cost of reducing multipole resolution.  We determine the 
midpoint of each band as the point where half the power in the band is below the midpoint and half above, and the endpoints of each band where the power falls to $e^{-1/2}$ of the peak power.  This step is 
useful for plotting the results, though we use the non-recombined angular power spectra in our $\chi^2$ fits.

\subsection{$\chi^2$ Fitting}
\label{Fitting}

After determining the angular power spectra of our data, we want to compare our observed results to the results predicted by a given theory, or alternatively to determine the theoretical 
parameters that best fit our data.  We do this by constructing a theoretical angular power spectrum from a prior 3D power spectrum that is dependent on cosmology.  We then project this 3D power spectrum down 
to two dimensions using Limber's approximation (\citealt{limber53,crocce10}) and the redshift distribution of the sample under consideration:

\begin{equation}
\label{EQ Limber Theory APS}
\mathcal{C}_{\ell}^T \approx \frac{2\pi}{\ell(\ell+1)}\int \phi^2(z) D^2(z) P(\frac{\ell + 1/2}{r(z)}) \frac{H(z)}{r^2(z)} b^2\ dz\
\end{equation}

\begin{equation}
\phi(z) = \frac{1}{\overline{G}}\ \frac{d\overline{G}}{dz}
\end{equation}
where $D(z)$ is the growth function (\citealt{carroll92}), $b$ is the bias, and $r$ and $\overline{g}$ are the comoving distance and number density respectively.

With the theoretical angular power spectrum, we then can compare with the observed angular power spectrum through $\chi^2$ fitting:

\begin{equation}
\chi^2(a_p) = \sum_{bb'} (\ln \mathcal{C}_b - \ln \mathcal{C}_b^T)\ \mathcal{C}_b F_{bb'} \mathcal{C}_{b'}\ (\ln \mathcal{C}_{b'} - \ln \mathcal{C}_{b'}^T)
\end{equation}
By minimizing this $\chi^2$, we can determine the 3D power spectrum that best fits the data and thus can constrain the parameters that produce that power spectrum.

\section{Fitting to Theory} 
\label{Fitting to Theory}

\subsection{Nonlinear Power Spectrum}
\label{Nonlinear Power Spectrum}

To improve the precision of our measured constraints on cosmological parameters over our previous fits, we implemented a theoretical angular power spectrum calculation using nonlinear 3D power spectra obtained 
by using the Code for Anisotropies in the Microwave Background (CAMB: \citealt{lewis00}).  Among its many functions, CAMB is capable of producing nonlinear 3D matter power spectra using HALOFIT 
(\citealt{smith03}).  HALOFIT generates nonlinear matter power spectra using the halo model to describe galaxy correlations (\citealt{seljak00,peacock00}).  The halo model suggests that the collisionless dark 
matter forms dark matter haloes within which baryons collapse to form some number of galaxies.  Thus, the matter power spectrum is separated into two regimes: the correlation of galaxies is determined by the 
correlation between different dark matter haloes on large scales, and the correlation between galaxies within the same haloes on small scales.  HALOFIT has determined empirical fitting functions for the power 
spectra of these two regimes by matching N-body simulations over a range of cosmological parameters.  The sum of the quasi-linear large scale term and the nonlinear small-scale term produces an overall 
nonlinear matter power spectrum.

By running CAMB and HALOFIT with high accuracy options enabled, we produce nonlinear matter power spectra with an accuracy of $\sim 0.2\%$ (\citealt{howlett12}).  However, this accuracy does assume a number of
factors, including: the ionization history of the Universe, the Hubble parameter, the dark energy equation of state parameter, and the initial power spectrum.  Smaller issues in these calculations include a
number of other parameters such as the primordial Helium fraction and effective number of neutrino species.  Where appropriate, we assume a flat cosmology and WMAP 7-year best fit results and leave other
parameters at the CAMB default settings, which correspond to the current best measurements of these parameters.  We generate 50 matter power spectra from $z = 0$ to $z = 1$ with $\Delta z = 0.02$ for each set 
of $\Omega_m$, $\Omega_b$, and bias $b$ that are fit, and proceed in the calculation from Equation \ref{EQ Limber Theory APS} as before.

\subsection{Nonlinear Fits}
\label{Nonlinear Fits}

Now that this calculation can be extended to nonlinear scales, we can determine cosmological parameters more precisely than in HBR12.  In addition to the bias and $\Omega_m$, we also chose to fit the baryon 
density, $\Omega_b$.  We also investigated fitting the spectral index, $n_s$, but found that generally the spectral index altered the angular power spectra similarly to $\Omega_m$, thus causing a degeneracy
in the parameter fits.  As a result, we chose not to fit this parameter.

\begin{figure}
  \begin{center}
    \includegraphics[width=0.5\textwidth]{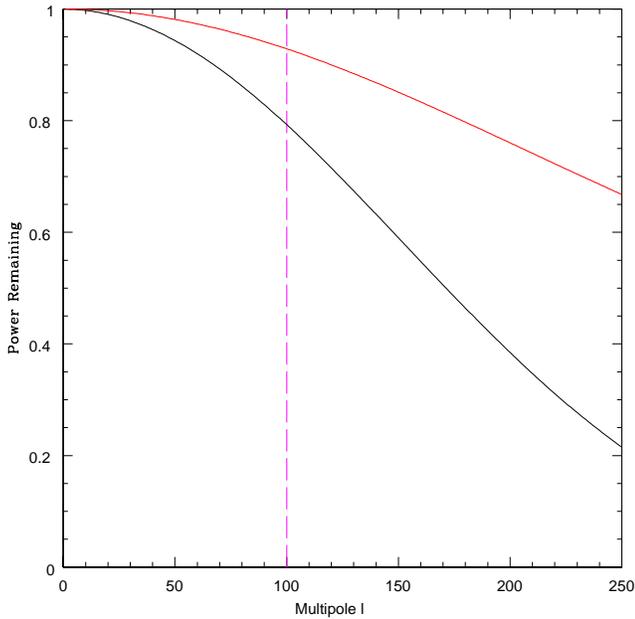}
    \caption{The pixel window function at HEALPix resolution 64 in black, demonstrating the percentage of power lost at each multipole for any angular power spectrum calculated at this pixelization scale.  In
      red, we integrate the pixel window function up to $\ell$ to show the cumulative power lost to that multipole.  We also show the maximum multipole we use in our $\chi^2$ fits at $\ell=100$, equivalent to
      twice the linear scale of the pixel.}
    \label{Pixel Window Function}
  \end{center}
\end{figure}

Before fitting, we must determine the maximum multipole to which we can fit the estimated and theoretical angular power spectra.  In some cases, we are limited by the signal-to-noise of the data, as we can't 
fit into the noise dominated region of our observed angular power spectra.  But in most samples, the signal-to-noise is sufficient for the full range of $\ell$ in our estimation and we are instead limited by 
the pixelization scale.  The pixelization process causes a loss of information on scales near the pixel size, but this is not a sharply defined boundary.  Instead, the pixelization discards 
steadily more information as you approach the pixel size.  The power lost due to pixelization is given by the pixel window functions, $w_{\ell}$, calculated and supplied by 
HEALPix\footnote{http://healpix.jpl.nasa.gov/html/intronode14.htm}, which is demonstrated in Figure \ref{Pixel Window Function} for resolution 64 (\textit{i.e.} HEALPix NSIDE parameter 64), the resolution of 
our volume-limited sample and subsample results.  We see that by $\ell \sim 175$, the pixelization supresses $50\%$ of the power in the angular power spectrum, 
even though the equivalent linear scale of a pixel is $\ell \sim 200$.  The pixel window functions relate the observed pixelated angular power spectrum $\mathcal{C}_{\ell}^{pix}$ and the underlying unpixelized 
angular power spectrum $\mathcal{C}_{\ell}^{unpix}$ by:
\begin{equation}
\mathcal{C}_{\ell}^{pix} = w_{\ell}^2 \mathcal{C}_{\ell}^{unpix}
\end{equation}
This has the effect of strongly supressing the theoretical angular power spectrum at high $\ell$, but also reducing power at all scales.  We have chosen to fit to a maximum multipole equivalent to twice the
pixel scale for all nonlinear theory fits, where the bandpower value is reduced by roughly 20\% by the pixel window function, but is still dominated by the unpixelized angular power spectrum.

Finally, we must be aware of a known systematic (\cite{tegmark02}) in this quadratic estimation method that causes an excess of power at the small scale (\textit{i.e.} high $\ell$) end of the angular power 
spectrum.  Typically, this only affects the last few bandpowers in our estimation; and to correct for this effect, we use the quadratic estimator to calculate out to scales equivalent to $\sim 80\%$ of the 
pixel size, and discard the extra bandpowers.  However, due to covariance between the bandpowers, some of this power is inevitably transferred onto small scales, complicating the exact calculation.

\section{Volume-Limited Sample} 
\label{Volume-Limited Sample}

For the entire SDSS DR7 galaxy sample, we are magnitude-limited, which biases the sample toward detecting intrinsically brighter galaxies at higher redshifts.  To improve on this, we construct a volume-limited 
sample out to redshift $z = 0.4$ to examine a complete sample that is relatively free from Malmquist bias.  This volume-limited sample was created by \cite{ross10}, selecting over 3.2 million DR7 galaxies with 
de-reddened r-band apparent magnitudes $m_r < 21$ to ensure galaxy completeness, r-band absolute magnitudes $M_r < -21.2$, and photometric redshifts $z < 0.4$.  This absolute magnitude cut is more stringent 
than the SDSS detection limit to account for differences in k-corrections between early- and late-type galaxies.  These data are also masked for seeing greater than 1.5 arcseconds, reddening greater than 0.2 
magnitudes, and areas of poor image quality, following the method outlined in HBR12.

\subsection{Volume-Limited Data} 
\label{Volume-Limited Data}

\begin{figure}
  \includegraphics[width=0.5\textwidth]{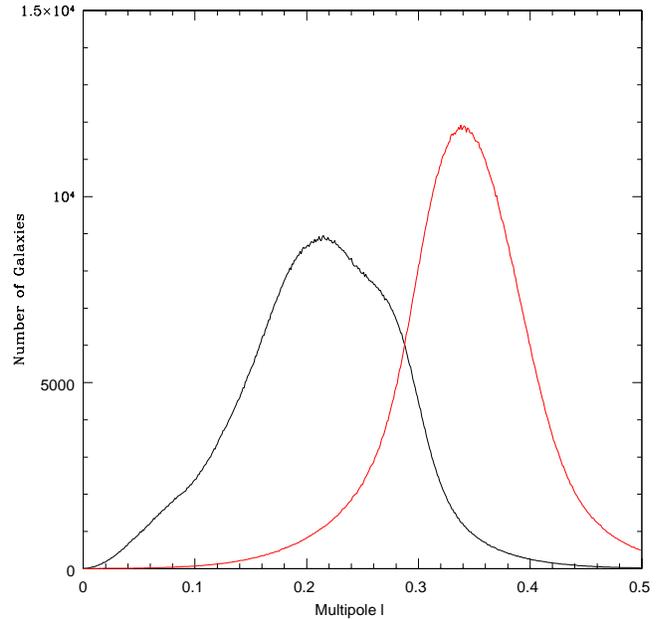}
  \caption{The redshift distributions of the redshift samples with the $z<0.3$ sample in black and $0.3<z<0.4$ in red.  These samples cover approximately equal cosmic volumes; but the high redshift 
    sample is more peaked than the low redshift sample.}
  \label{Redshift Distribution Graph}
\end{figure}

This volume-limited data set contains 3.2 million galaxies between redshift $0.0 < z < 0.4$.  We have additionally separated this volume-limited sample into two mutually exclusive redshift slices of 
approximately equal cosmic volume from $z < 0.3$ and $0.3 < z < 0.4$ to examine the possible evolution of the angular power spectrum with redshift and possible variation of the fit constrained cosmological 
parameters.  The redshift distributions of the two redshift slices is shown in Figure \ref{Redshift Distribution Graph}.

\subsection{Volume-Limited Angular Power Spectrum} 
\label{Volume-Limited Angular Power Spectrum}

\begin{figure}
  \begin{center}
    \includegraphics[width=0.5\textwidth]{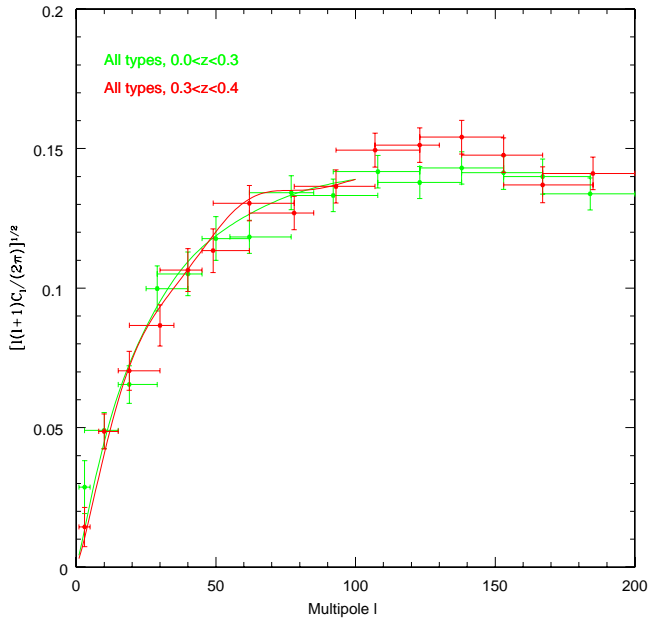}
    \caption{The best fit theoretical angular power spectrum to the samples split up by redshift shell with the $z<0.3$ sample in green and the $0.3<z<0.4$ sample in red.  Also apparent in the theoretical
    curves for the high redshift sample, plotted for $\ell < 100$,  are the wiggles from baryon acoustic oscillations.  The BAOs are smoothed over in the low redshift sample due to the larger redshift range.}
    \label{Theory All Types}
  \end{center}
\end{figure}

\begin{figure}
  \includegraphics[width=0.5\textwidth]{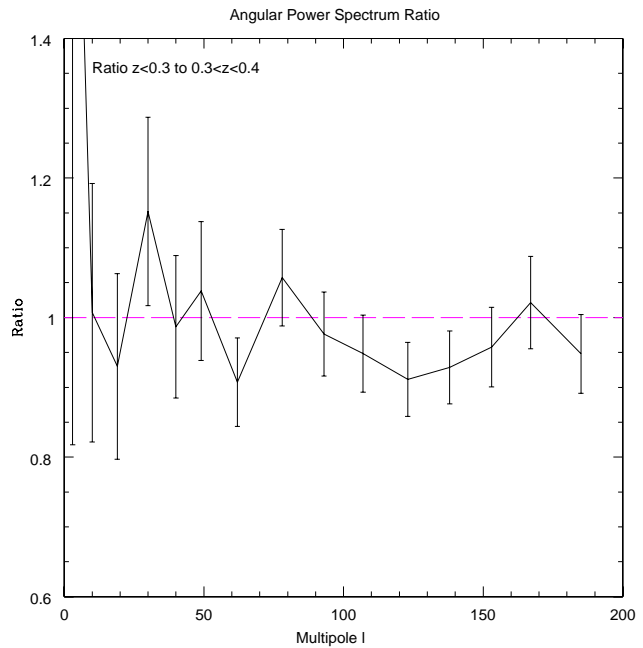}
  \caption{The ratio of the angular power spectra of the redshift samples $z<0.3$ and $0.3<z<0.4$.  There is a slight dip for $100<\ell<150$ where the high redshift sample has more power, but generally
  the results are consistent with a ratio of one.}
  \label{Volume-Limited Results Ratio Graph}
\end{figure}

We have calculated the angular power spectra of our low redshift $z < 0.3$ and high redshift $0.3 < z < 0.4$ samples and show them in Figure 
\ref{Theory All Types}.  We can see that generally these two angular power spectra look roughly indistinguishable, though perhaps there is some variation at multipoles $100 < \ell < 150$.  To examine this 
closer, we have taken the ratio of these angular power spectra, which is shown in black in Figure \ref{Volume-Limited Results Ratio Graph}.  We see that the ratio is fairly consistent with a value of one over 
the range of theoretical fitting, $\ell<100$, which implies that we cannot confidently detect any significant evolution at these scales.  Beyond the range that we fit a theoretical angular power spectrum to, 
$100<\ell<150$, we do see a small deviation from unity, slightly outside the $1\sigma$ error bars, which is not reflected in our fit values.  However, beyond $\ell=150$, we see that the values return to 
consistency with unity.

Recall that although our samples are of galaxies with measured photometric redshifts $z < 0.3$ and $0.3 < z < 0.4$ and are therefore mutually exclusive, each of these galaxies has a photometric redshift error
associated with them.  We assume a Gaussian probability distribution function for each galaxy with mean equal to the measured photometric redshift and standard deviation of the measured photometric redshift 
error.  As in \cite{ross10}, we take 10 samples from the PDF of each galaxy, which are then weighted by luminosity function (\cite{montero-dorta09}) and volume constraints to account for the lower likelihood 
of both the galaxy being bright and residing in a smaller cosmic volume (\textit{i.e.} lower redshift).  The galaxy redshift samples are normalized so that each galaxy contributes equally to the estimated 
``true'' redshift distribution, which is shown in Figure \ref{Redshift Distribution Graph} for the redshift subsamples of the volume-limited data set.  Since these redshift distributions overlap, we expect 
covariance between the measurements in each sample.  Therefore, we expect some similarities in the features of the angular power spectra.

\subsection{Fitting Volume-Limited Results} 
\label{Fitting Volume-Limited Results}

We have also produced fits out to a maximum of $\ell = 100$ (twice the pixel size) of the nonlinear theoretical angular power spectra to the volume-limited sample and subsamples.  Due to the power associated 
with a high stellar density in the first bandpower of the late-type galaxy angular power spectra (see, \textit{e.g.}, Figure \ref{Galaxy Morphology Results Graph}), discussed in Section 
\ref{Late-Type Large Scale Power}, we have not included the first bandpower in these fits.  To be consistent, we have excluded the first bandpower from all fits derived from the volume-limited sample, though 
the effect on the non-late-type galaxy samples was small, changing the best-fit parameters by $\sim 0.001$.  These fits are generally consistent with the results in Section \ref{Large Multipole Sample}, and the 
WMAP 7-year results and show no strong evidence of evolution with redshift.
These best fit theoretical spectra shown in Figure \ref{Theory All Types}.  Of interest to note is that the baryon acoustic oscillations (BAOs) (\citealt{eisenstein05,seo12}) are clearly visible in the 
$0.3 < z < 0.4$ sample due to its narrow redshift range, whereas the BAOs are smoothed over in the larger redshift ranges of the $z < 0.3$ and $z < 0.4$ samples.  The best fit values of bias and cosmological 
parameters for these samples are given in Table \ref{Table Volume Limited Best Fits}.

\begin{table}
  \begin{center}
    \begin{tabular}{| c | c | c | c | c |}
      \hline
      Redshift & Type & Bias & $\Omega_m$ & $\Omega_b$ \\
      \hline                                                                              
      Full-z & All & $1.545 \pm 0.057$ & $0.282 \pm 0.026$ & $0.041 \pm 0.020$ \\
      Full-z & Early & $1.727 \pm 0.065$ & $0.275 \pm 0.025$ & $0.033 \pm 0.019$ \\
      Full-z & Late & $1.256 \pm 0.051$ & $0.262 \pm 0.030$ & $0.039 \pm 0.023$ \\
      \hline
      Low-z & All & $1.421 \pm 0.054$ & $0.300 \pm 0.036$ & $0.026 \pm 0.024$ \\
      Low-z & Early & $1.666 \pm 0.064$ & $0.351 \pm 0.038$ & $0.034 \pm 0.022$ \\
      Low-z & Late & $1.172 \pm 0.050$ & $0.279 \pm 0.041$ & $0.032 \pm 0.032$ \\
      \hline
      High-z & All & $1.515 \pm 0.057$ & $0.234 \pm 0.019$ & $0.036 \pm 0.014$ \\
      High-z & Early & $1.634 \pm 0.062$ & $0.239 \pm 0.019$ & $0.033 \pm 0.014$ \\
      High-z & Late & $1.363 \pm 0.062$ & $0.244 \pm 0.024$ & $0.036 \pm 0.017$ \\
      \hline
    \end{tabular}
    \caption{The best fit biases and cosmological parameters for the volume-limited sample and subsamples.}
    \label{Table Volume Limited Best Fits}
  \end{center}
\end{table}

\section{Galaxy Morphology} 
\label{Galaxy Morphology}

We also separate the volume-limited sample into early- and late-type galaxies, and, in addition to clearly showing the stronger clustering of early-type galaxies, we are able to effectively determine the 
relative linear bias between these two morphological types in Section \ref{Galaxy Morphology Results}.  These samples are created by separating early- and late-type galaxies based on the \texttt{pztype} 
parameter provided by the SDSS.  The \texttt{pztype} parameter is an estimate of the spectral type of the galaxy calcuated in the photometric redshift pipeline (\citealt{abazajian09}), and we classify 
galaxies with a \texttt{pztype} value of less than $0.1$ as early-type and those with \texttt{pztype} greater than $0.1$ as late-type (\citealt{ross10}).  

In Section \ref{Combining Redshift and Galaxy Morphology Cuts}, we combine the redshift and galaxy type cuts together to produce four more subsamples that allow us to examine the possible redshift 
evolution of the angular power spectrum of early- and late-type galaxies.

\subsection{Galaxy Morphology Results} 
\label{Galaxy Morphology Results}

As early-type galaxies are believed to preferentially form in high density environments, we wish to explore evolution in the angular power spectra of galaxies split by galaxy type.  In our volume-limited
sample, we can compare the angular power spectra of early-type galaxies, which tend to be larger and brighter, to late-type galaxies.  This is important since by using a volume-limited sample we avoid 
Malmquist bias, which would result from selectively detecting only brighter galaxies in a magnitude-limited sample, which have more power at all scales.

\begin{figure}
  \includegraphics[width=0.5\textwidth]{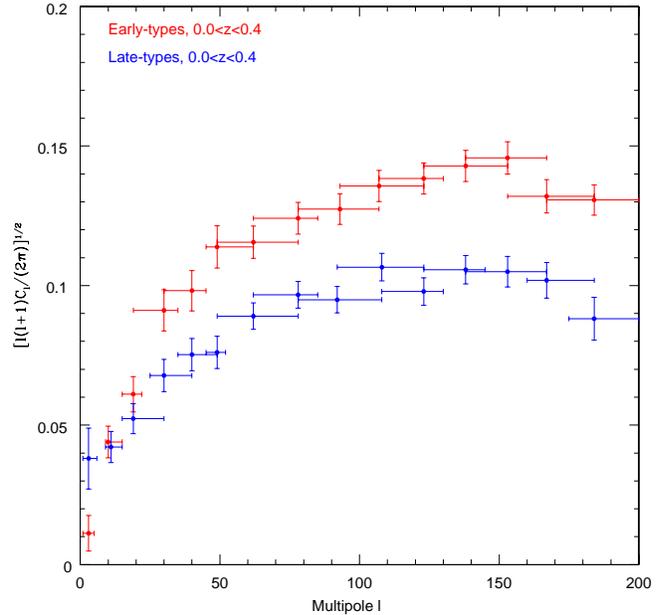}
  \caption{The angular power spectra of the early- and late-type galaxies in the volume-limited sample.}
  \label{Galaxy Morphology Results Graph}
\end{figure}

\begin{figure}
  \includegraphics[width=0.5\textwidth]{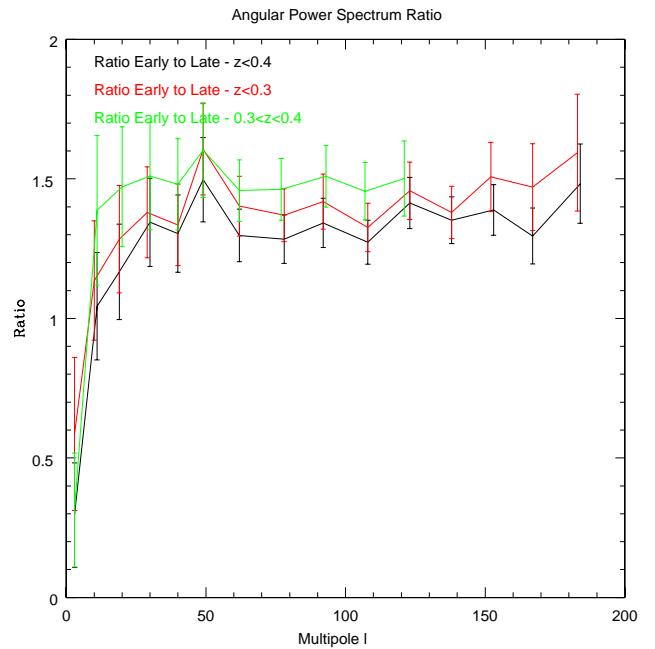}
  \caption{The ratio of the angular power spectra of the early- to late-type galaxies in the volume-limited sample.}
  \label{Galaxy Morphology and Redshift Results Ratio Graph}
\end{figure}

We also want to estimate the linear bias of early-type and late-type galaxies, and calculate the relative bias between them.  The results of our galaxy morphology angular power spectra are given in Figure
\ref{Galaxy Morphology Results Graph}.  Since the relative linear bias is roughly the ratio between these two angular power spectra, we can easily estimate the relative bias between early- and late-type
galaxies by examining the ratio in black in Figure \ref{Galaxy Morphology and Redshift Results Ratio Graph}.  We can see that the relative bias is remarkably consistent across all but the very largest scales
down to $\sim 1$ degree, and the bias of early-type galaxies is roughly $30-40\%$ greater than the bias of late-type galaxies.  
The best fit values of bias and cosmological parameters for these samples are presented in Table \ref{Table Volume Limited Best Fits}.

\subsection{Combining Redshift and Galaxy Morphology Cuts} 
\label{Combining Redshift and Galaxy Morphology Cuts}

\begin{figure}
  \begin{center}
    \includegraphics[width=0.5\textwidth]{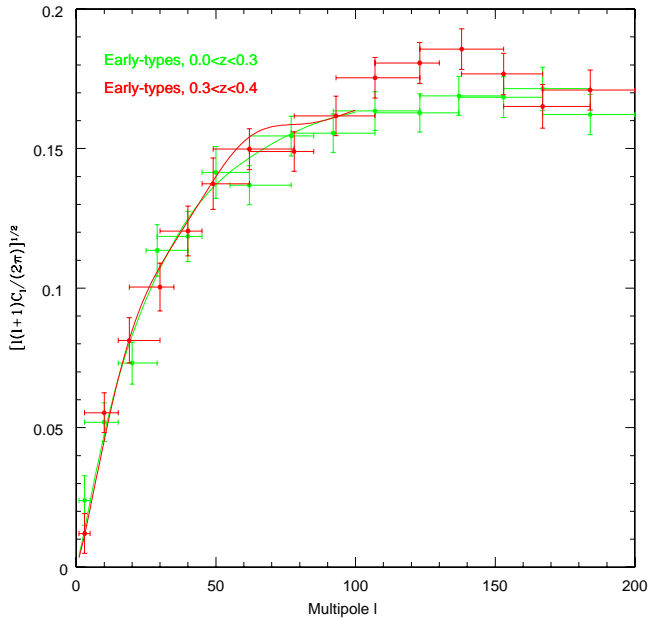}
    \caption{The best fit theoretical angular power spectrum to the early-type samples split up by redshift shell with the $z<0.3$ sample in green and the $0.3<z<0.4$ sample in red.  Also apparent in the
    theoretical curves for the high redshift sample are the wiggles from baryon acoustic oscillations.  The BAOs are smoothed over in the low redshift sample due to the larger redshift range.}
    \label{Theory Early Types}
  \end{center}
\end{figure}

\begin{figure}
  \begin{center}
    \includegraphics[width=0.5\textwidth]{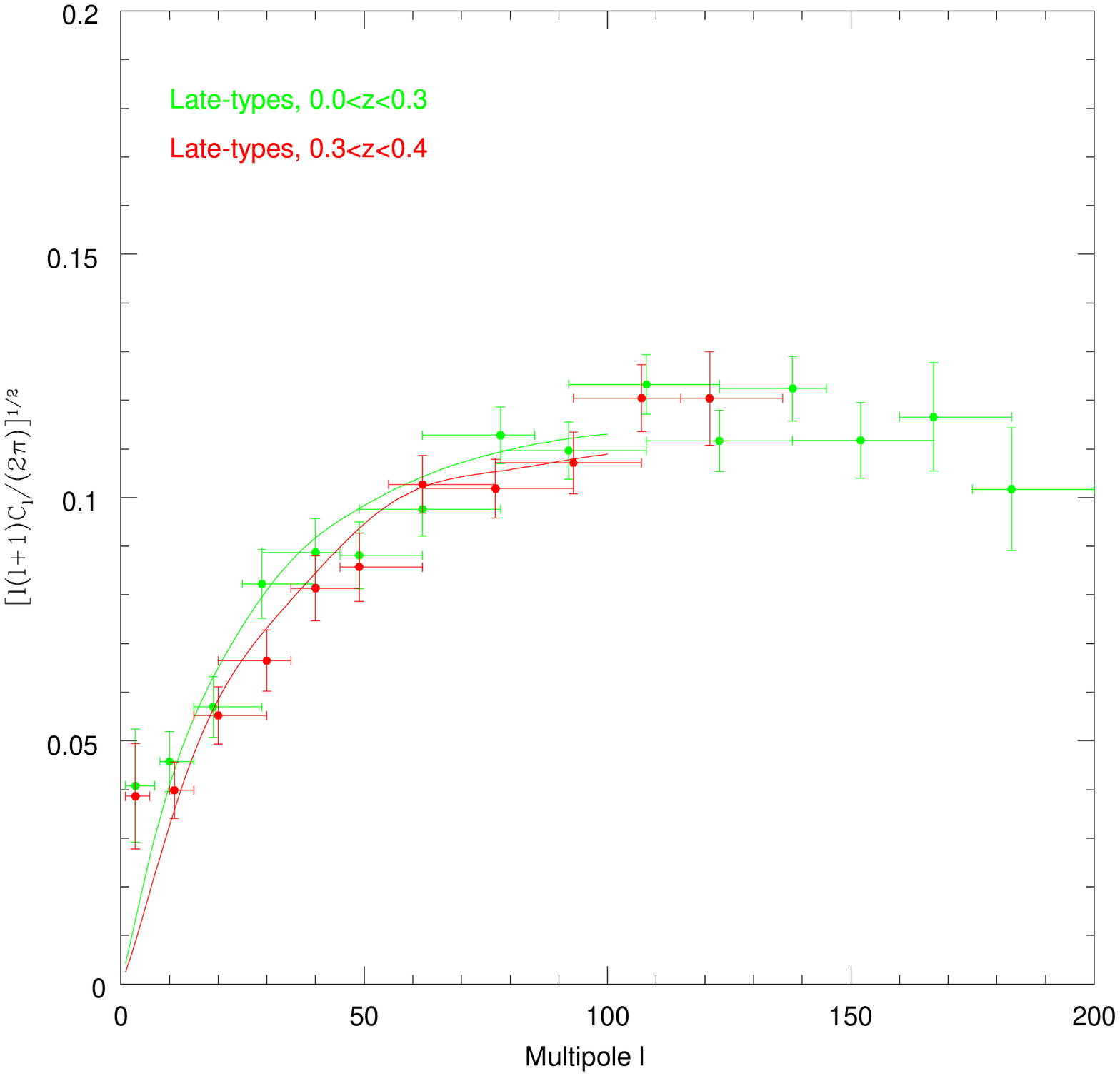}
    \caption{The best fit theoretical angular power spectrum to the late-type samples split up by redshift shell with the $z<0.3$ sample in green and the $0.3<z<0.4$ sample in red.  Also apparent in the
    theoretical curves for the high redshift sample are the wiggles from baryon acoustic oscillations.  The BAOs are smoothed over in the low redshift sample due to the larger redshift range.}
    \label{Theory Late Types}
  \end{center}
\end{figure}

\begin{figure}
  \includegraphics[width=0.5\textwidth]{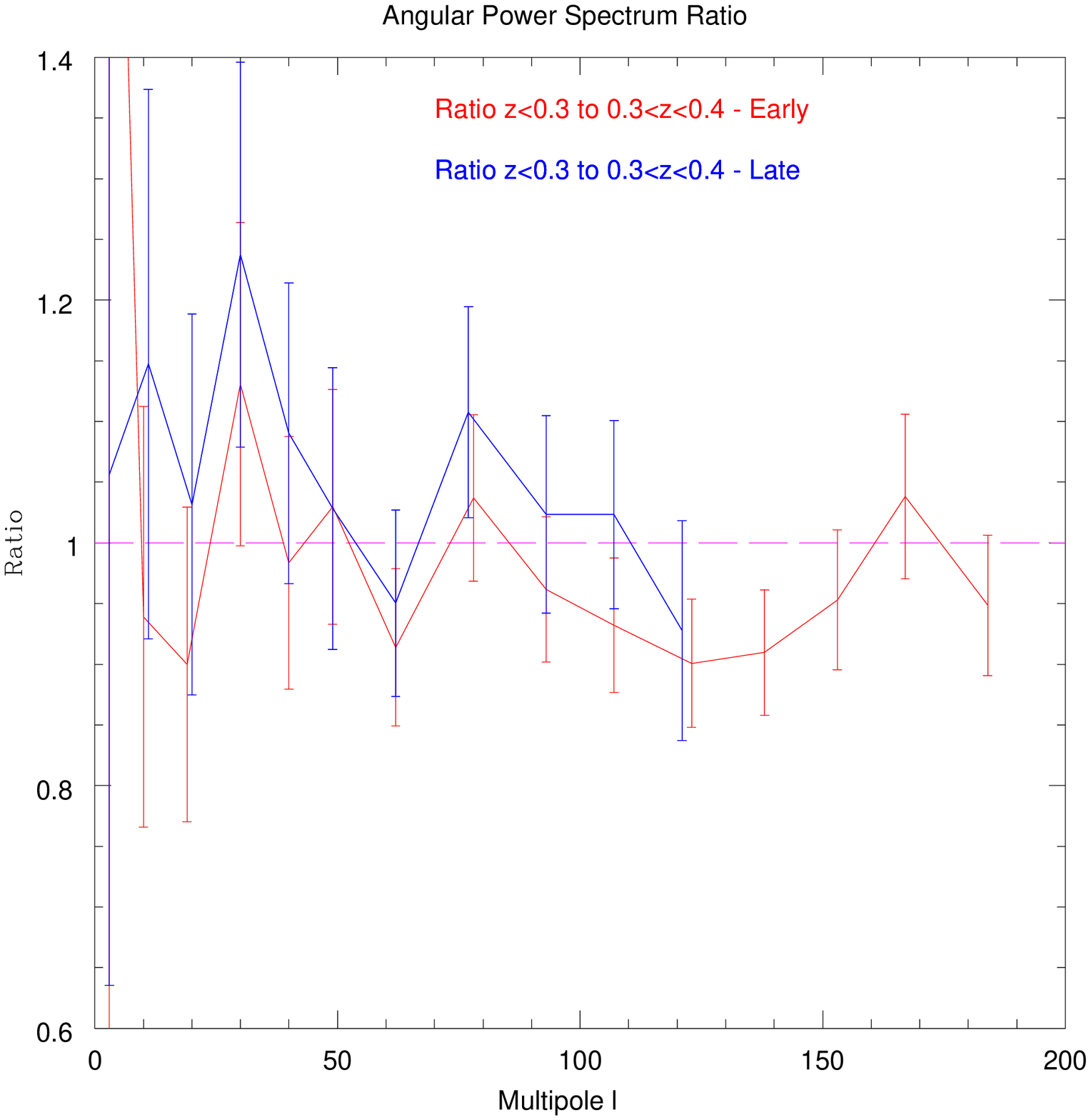}
  \caption{The ratio of the angular power spectra of the redshift samples $z<0.3$ and $0.3<z<0.4$, separated by type.  The late-type high redshift sample is signal-to-noise limited beyond $\ell=120$ and
  we truncate the angular power spectrum for that sample.  We see a slightly more pronounced dip for $100<\ell<150$ in the early-type galaxies compared to Figure \ref{Volume-Limited Results Ratio Graph}, but
  we still can't strongly conclude that there is evidence of significant evolution in redshift.}
  \label{Volume-Limited Results by Type Ratio Graph}
\end{figure}

The results of our estimation of the angular power spectra of the different redshift slices for early- and late-type galaxies are given in Figures \ref{Theory Early Types} and \ref{Theory Late Types}.  We see, 
as in Section \ref{Volume-Limited Angular Power Spectrum}, no clear evidence of evolution between these two samples in different redshift ranges.  Note that the high redshift late-type results are also 
signal-to-noise limited around $\ell \sim 120$, we therefore truncate that angular power spectrum.  The ratio of angular power spectra of these samples are plotted along with the results for all types above in 
Figure \ref{Volume-Limited Results by Type Ratio Graph}, and are also generally consistent with a value of one suggesting no evidence of evolution.

Finally, we want to look for the possible evolution of early- and late-type galaxies between our two redshift samples.  By comparing the linear biases of the early- and late-type galaxies, we can also determine 
the relative linear bias of these galaxy types.  We find that the relative bias $b_e/b_l = 1.375 \pm 0.076$ for our entire $z<0.4$ volume-limited sample, while $b_e/b_l = 1.421 \pm 0.083$ for $z<0.3$ and 
$b_e/b_l = 1.200 \pm 0.071$ for $0.3<z<0.4$.  These are fairly consistent across all samples, and show no evidence of significant scale dependence on these large scales.

\section{Large Multipoles}
\label{Large Multipoles}

We would like to determine the angular power spectrum to the highest multipoles (and thus smallest scales) allowed by the SDSS DR7 galaxy data, and to examine these small scales requires a high resolution 
pixelization.  However, we are computationally limited by the $O(n_p^3)$ scaling dependence of the quadratic estimator (\citealt{borrill99}), where $n_p$ is the number of pixels in the map.  By using large 
shared-memory supercomputers, this limits the analysis to maps with $n_p \sim 10^4$; thus, in order to go to higher resolution and multipoles in the absence of better computational methods, we are forced to 
constrain our analysis to smaller areas than the full SDSS DR7.  Performing this calculation with a smaller area necessitates using larger bands as the band width is inversely proportional to the smallest 
linear dimension of the area under consideration (\citealt{peebles80}).  Therefore, obtaining the angular power spectrum at high multipoles involves sacrificing band resolution and survey area, which means 
losing information about the angular power spectrum at all scales.  However, this sacrifice is offset by being able to constrain the angular power spectrum at smaller angular scales where cosmological 
parameters have a greater effect, ideally this will allow tighter constraints on those parameters.

\subsection{Large Multipole Results}
\label{Large Multipole Results}

To extend the angular power spectrum out to high multipoles, we need high signal-to-noise and therefore we use the full magnitude-limited data of HBR12 instead of the volume-limited subset.  This includes all 
galaxies with r-band magnitudes in the range 18--21 that have photometric redshifts and associated errors.

\begin{figure}
  \includegraphics[width=0.5\textwidth]{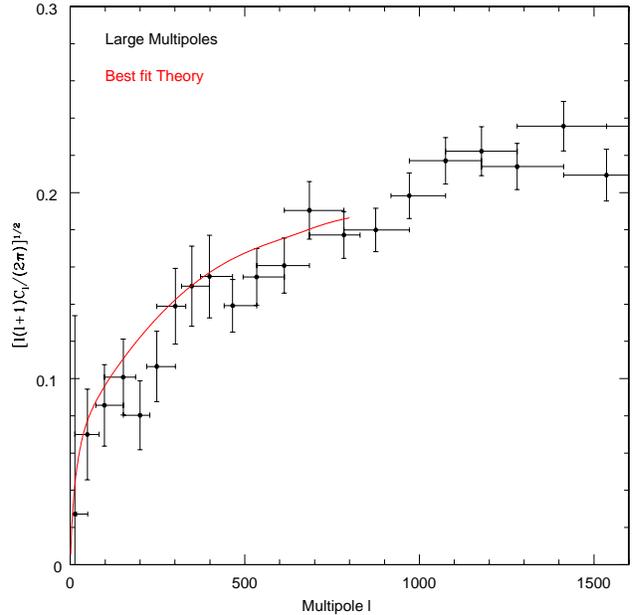}
  \caption{In black, we show the angular power spectrum out to $\ell = 1600$ of an $\sim 53.7$ square degree area of the SDSS DR7 near the North Galactic Pole, at an eighth of the linear scale of the full area 
    results.  The best fit theoretical angular power spectrum to the large multipole, high resolution sample out to $\ell=800$, equivalent to twice the linear scale of the HEALPix resolution 512 pixels, is 
    shown in red.  Note that the points in black have had bandpowers combined together for higher signal-to-noise presentation, the best fit is determined from the non-combined angular power spectrum.}
  \label{Large Multipole Results Graph}
\end{figure}

When restricting our analysis to a smaller area, we are at least free to choose the particular area we examine.  We select an area of low reddening near the North Galactic Pole, and limit our analysis to the
$\sim 53.7$ square degree area corresponding to nested pixel number 162 at HEALPix resolution 8.  For the calculation, we use $\sim 0.013$ square degree pixels at HEALPix resolution 512, which allows us to 
compute bands of width $25 \ell$ out to $\ell \sim 1600$, the scale roughly equivalent to the pixel size.  This sample has a mean redshift of $z = 0.251$.  We present the results from this measurement in Figure 
\ref{Large Multipole Results Graph}.

\subsection{Fitting Large Multipole Sample}
\label{Large Multipole Sample}

Using the nonlinear power spectra produced by CAMB, we have fit the results of our large multipole sample for the parameters $\Omega_m$, $\Omega_b$, and bias.  We have fit out to a maximum of $\ell = 800$ or 
twice the pixel size to avoid signal lost by the pixelization.  We find that the best fit $\Omega_m = 0.267 \pm 0.038$, $\Omega_b = 0.045 \pm 0.012$, and $b = 1.075 \pm 0.056$, and show this fit in Figure 
\ref{Large Multipole Results Graph}.

\section{Discussion}
\label{Discussion}

\subsection{Late-Type Large Scale Power}
\label{Late-Type Large Scale Power}

In Figures \ref{Galaxy Morphology Results Graph} and \ref{Galaxy Morphology and Redshift Results Ratio Graph}, it is clear that the power in our first bandpower for late-type galaxies is unusually high, and
this is consistent across all late-type samples.  The smallest scale probed by this range of $\ell$ is over 30 degrees, a great deal larger than where we expect significant structure to exist.  This suggests
that despite our masking process correcting for seeing, reddening, and poor observing quality, there may be a systematic that preferentially affects late-type galaxies on large scales.  We have examined our
late-type sample in more detail, and have found that the average density of late-type galaxies drops at either end (in lambda, the survey longitude) of the SDSS observing footprint, nearer to 
the Galactic plane.
While we already discussed in HBR12 that the overall galaxy overdensity was consistent with zero at all Galactic latitudes, the late-type sample is more significantly affected by this large scale systematic 
effect and this is apparent in the first bandpower of these angular power spectra.

As this effect occurs closer to the Galactic plane, systematics that could reduce late-type galaxy counts include stellar obscuration of background galaxies, variable star-galaxy separation efficiency, and an
insufficiently strict reddening mask cut.  However, closer to the Galactic plane is also higher in stripe longitude lambda, so it is possible that variable sky brightness could influence the observed galaxy
density.  

\begin{figure}
  \includegraphics[width=0.5\textwidth]{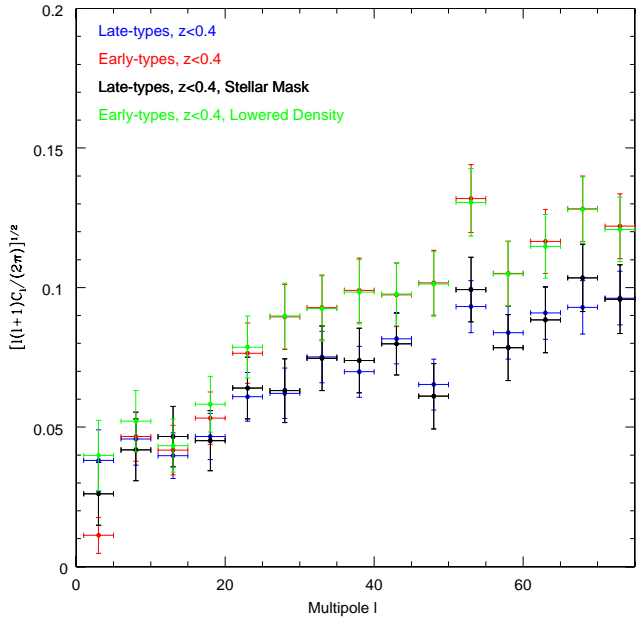}
  \caption{The angular power spectrum of late-type galaxies, in blue, compared with the same sample masked for high stellar overdensities in black.  We see that the power in the first band is significantly 
    reduced, but still higher than expected.  Also in this plot, the angular power spectrum of early-type galaxies, in red, compared the the same sample with densities artificially lowered in high stellar 
    density pixels in green.  This demonstrates that anti-correlation in star and galaxy densities can result in this large scale power, leaving the rest of the angular power spectrum largely unaffected.}
  \label{Lowered Density APS}
\end{figure}

In order to identify the issue causing this underdensity of late-type galaxies, we need to first determine how to distinguish these various, nearly degenerate, possible causes.  
First, we extended our algorithm to calculate a cross-correlation angular power spectrum.  By using this new code, we computed the cross power spectrum between late-type galaxies and stars, early-type
galaxies and stars, and late-type galaxies and reddening.  From these measurements, we find that only the late-type/star cross power spectrum shows significant anti-correlation at small $\ell$.  This evidence
implies high stellar densities are associated with the slight drop in density of late-type galaxies, and suggests that either stellar obscuration or star-galaxy separation could be responsible.

To test the possibility that stars are interfering with late-type galaxy densities, we pixelated the stars in the SDSS data in the same manner as galaxies and used these to apply high stellar density masks to 
our late-type galaxy sample.  We found that stringent cuts on high stellar density pixels partially reduced, but could not eliminate, this large scale power, as shown in black in Figure 
\ref{Lowered Density APS}.

We also attempted to replicate this effect in the early-type galaxies, which do not show this large scale power.  We used the above stellar density masks and artificially lowered the early-type galaxy density 
in those pixels that have a higher density of stars.  Due to the increased density of stars near the Galactic plane, this affects nearly all pixels at low Galactic latitudes and high lambda.  However, we see 
in green in Figure \ref{Lowered Density APS} that this does raise the large scale power in the first bandpower in the same manner as the late-type galaxy samples, suggesting a correlation between the 
increased density of stars and the lower than expected late-type galaxy density.

\begin{figure}
  \includegraphics[width=0.5\textwidth]{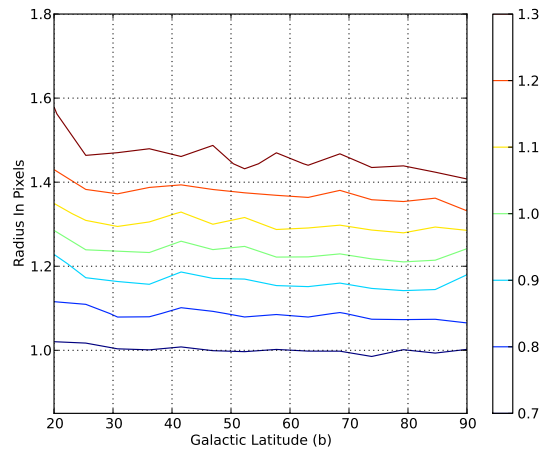}
  \caption{The ratio of the late-type galaxy radius distribution to the early-type galaxy radius distribution as a function of Galactic latitude, b.  We see that the distribution of galaxy radii change 
    significantly at low latitudes.}
  \label{Radius Distribution Comparison}
\end{figure}

With these results in mind, we next investigated how the properties of the late-type galaxy distribution changes as a function of Galactic latitude.  By comparing the measured size of SDSS galaxies using the
Petrosian radius containing half the Petrosian flux (\citealt{petrosian76}), we generated normalized 2D histograms of of the late- and early-type galaxy radii as a function of Galactic latitude.  We 
plot the ratio of these normalized distributions in Figure \ref{Radius Distribution Comparison}, which shows that the galaxy radius distribution changes at low Galactic latitudes.  The late-type galaxies at low 
latitudes are on average larger than at high latitudes, which when combined with the overall reduction in low latitude late-type galaxy density suggests that our samples have undercounted small late-type 
galaxies in high stellar density fields.  Our interpretation of these results is that small late-type galaxies in the SDSS are being misclassified as stars due to deblending issues in crowded fields.

To ensure that this systematic does not affect our $\chi^2$ fits for cosmological parameters and bias, we have not included the first bandpower of the angular power spectra derived from our volume-limited 
sample in the fitting process.  We have excluded the first bandpower for all volume-limited sample fits for consistency. 

\subsection{Comparison with Previous Results}
\label{Comparison with Previous Results}

The constraints provided by our fits to nonlinear theoretical matter power spectra for our large multipole and volume-limited sample and subsamples are much improved over the linear fits used in HBR12.  We find 
that our results are generally consistent with an $\Omega_m \approx 0.27$ with errors on the order of $0.03$, which is typical of galaxy angular power spectra (\citealt{blake07}).  This agrees well with the 
WMAP7 CMB results (\citealt{larson11}) of $\Omega_m = 0.267 \pm 0.026$ as well as the combination of SDSS DR8 luminous galaxy angular power spectrum results with WMAP7 and supernova data 
$\Omega_m = 0.267 \pm 0.0163$ (\citealt{ho12}).

We find $\Omega_b \approx 0.03$ with errors generally about $0.02$ in our volume-limited and large multipole samples, which is consistent with the $\Omega_m = 0.0449 \pm 0.0028$ constraints produced by WMAP7.
The errors on our measurements are an order of magnitude larger than WMAP7, however, as our samples are less sensitive to this parameter, largely due to uncertainties in the photometric redshifts, and this is
similar in other galaxy angular power spectra results (\citealt{thomas10}).

Measurements of the bias parameter vary with the sample under consideration.  As we've shown in Section \ref{Galaxy Morphology Results}, bias is strongly type dependent, so the ratio of early- and
late-type galaxies in a sample has a profound effect on the bias.  Even the relative bias between early- and late-type galaxies has wide variations from a relative bias of $1.2 \pm 0.15$ (\citealt{willmer98})
to $\sim 1.75$ (\citealt{ross06}), and is partially dependent on the cut used to separate the different galaxy types.

With our assumptions of a flat Universe, and various properties of the initial angular power spectrum and reionization, these results imply that the mass-energy content in the Universe is not dominated by mass,
but by another form of energy, believed to be dark energy.  Even the mass in the Universe is not primarily normal, baryonic matter, but collisionless dark matter.  The galaxies that we observe make up just a
few percent of the mass-energy, but these galaxies trace the underlying dark matter distribution with a type-dependent bias describing how clustered galaxies are compared to the underlying dark matter.  It is
worth noting that these implications are quite consistent with the WMAP7 results, despite relying on an entirely independent measurement process and data set during a completely different cosmic epoch.

\subsection{Future Work}
\label{Future Work}

This technique can be easily applied to other surveys, such as the Dark Energy Survey (DES).  The DES is scheduled to begin taking data by the end of 2012, but even before that data 
is available, simulations of the DES data can be analyzed with our angular power spectrum estimation code similar to what was done in T02 with the SDSS Early Data Release.  To support this and similar 
efforts, we have made our parallelized estimation code publicly available\footnote{We have made these codes freely available at $http$:$//lcdm.astro.illinois.edu/code/apscode.html$}.

We also can see the evidence of baryon acoustic oscillations in the angular power spectrum for narrow redshift slices, as in the theoretical angular power spectra in Figures \ref{Theory All Types},
\ref{Theory Early Types}, and \ref{Theory Late Types}.  Though the variations in the angular power spectra from BAOs is generally smaller than our error bars for these samples, detection of the BAOs in SDSS
galaxy angular power spectra in the $30 < \ell < 300$ range is an area of current research (\citealt{seo12}).  Indeed, the Baryon Oscillation Spectroscopic Survey (BOSS: \citealt{eisenstein11}) of SDSS-III
aims to explore the BAO signal through examining correlations of Luminous Red Galaxies (LRGs) and quasars.

A closely related measurement that can be done is the cross-power spectrum estimation.  The method discussed in Section \ref{Quadratic Estimation} has referred only to auto-correlation, that is correlating 
galaxy overdensity with itself; but Equation \ref{EQ covariance matrix} can be extended to calculate the cross-correlation, for example between quasars and LRGs, by taking the outer product of two different 
data sets rather than of the same data set.  This will not only provide interesting science, such as constraints on local primordial non-Gaussianity (\citealt{slosar08}), but can also be a test of systematics 
by calculating the cross-correlation of two samples that should be uncorrelated, as we have seen in Section \ref{Late-Type Large Scale Power}. 

The advances in general purpose graphics processors and their incorporation into modern supercomputers suggests it is time to revisit the possibility of accelerating this computation on these machines.  The 
computationally intensive parts of this calculation are the signal matrix construction and the matrix multiplications and inversions required by the quadratic estimator, and we have previously found that these 
operations perform quite well on graphics processors (\citealt{hayes07}).  By accelerating these calculations, it might be possible to improve the resolution while not sacrificing survey area, permitting 
angular power spectrum measurements with both high bandpower resolution and results extending to smaller scales.  This could improve cosmological parameters constraints from these galaxy surveys.

\section{Conclusions} 
\label{Conclusions}

We have used a quadratic estimator with KL-compression to determine the angular power spectrum of the SDSS DR7.  We applied this technique to a $\sim 53.7$ square degree area near the North Galactic Pole to 
determine the angular power spectrum of the SDSS DR7 main galaxy sample out to high multipoles ($\ell = 800$).  By utilizing CAMB to produce nonlinear power spectra, we created theoretical nonlinear angular 
power spectra by projecting these CAMB spectra to two dimensions using the photometric redshifts of the SDSS data.  By minimizing the $\chi^2$ fits between these theoretical angular power spectra and our 
measured SDSS angular power spectra, we find that the best-fit cosmological parameters to this SDSS DR7 main galaxy sample are $\Omega_b = 0.045 \pm 0.012$ and $\Omega_m = 0.267 \pm 0.038$, with a linear bias of 
$b = 1.075 \pm 0.056$.  

Using the same process on a volume-limited sample covering the Northern contiguous SDSS footprint out to $\ell = 100$, we determine $\Omega_b = 0.041 \pm 0.020$, $\Omega_m = 0.282 \pm 0.026$, and 
$b = 1.545 \pm 0.057$, while the linear bias varies from $b = 1.515 \pm 0.057$ in galaxies from $0.3 < z < 0.4$ to $b = 1.421 \pm 0.054$ in the range $0.0 < z < 0.3$.  Splitting the volume-limited sample by 
type, we calculate the relative bias $b_e/b_l = 1.375 \pm 0.076$, and combining this information with the photometric redshift slices, we see that the relative bias varies from $b_e/b_l = 1.200 \pm 0.071$ in 
the $0.3 < z < 0.4$ range to $b_e/b_l = 1.421 \pm 0.083$ when $0.0 < z < 0.3$.

Finally, our examination of the angular power spectrum by galaxy type has shown unexpected large scale power in late-type galaxies.  This suggests that late-type galaxies near the Galactic plane are 
undercounted, and this is associated with areas of high stellar density.  Our analysis accounted for this by excluding the largest scales from the fits to determine cosmological parameters.

\section*{Acknowledgements}
\label{Acknowledgements}

The authors would like to thank Ben Wandelt for valuable discussion and advice used in this paper, and the referee for suggestions that have improved this manuscript.

This research was supported in part by the National Science Foundation through XSEDE resources provided by Pittsburgh Supercomputing Center's 4,096 core SGI UV 1000 (Blacklight).

Some of the results in this paper have been derived using the HEALPix \citep{gorski05} package.

Funding for the SDSS and SDSS-II has been provided by the Alfred P. Sloan Foundation, the Participating Institutions, the National Science Foundation, the U.S. Department of Energy, the National Aeronautics
and Space Administration, the Japanese Monbukagakusho, the Max Planck Society, and the Higher Education Funding Council for England. The SDSS Web Site is $http$:$//www.sdss.org/$.

The SDSS is managed by the Astrophysical Research Consortium for the Participating Institutions. The Participating Institutions are the American Museum of Natural History, Astrophysical Institute Potsdam,
University of Basel, University of Cambridge, Case Western Reserve University, University of Chicago, Drexel University, Fermilab, the Institute for Advanced Study, the Japan Participation Group, Johns Hopkins
University, the Joint Institute for Nuclear Astrophysics, the Kavli Institute for Particle Astrophysics and Cosmology, the Korean Scientist Group, the Chinese Academy of Sciences (LAMOST), Los Alamos National
Laboratory, the Max-Planck-Institute for Astronomy (MPIA), the Max-Planck-Institute for Astrophysics (MPA), New Mexico State University, Ohio State University, University of Pittsburgh, University of
Portsmouth, Princeton University, the United States Naval Observatory, and the University of Washington.

\end{document}